\documentclass[a4paper]{article}
\usepackage{subfigure}
\usepackage{icrc2013}
\usepackage[english]{babel}
\title{Propagation of UHECRs in cosmological backgrounds:\\
       some results from {\it SimProp}.}

\shorttitle{{\it SimProp} Propagation Code}

\authors{
R. Aloisio$^{1,2}$,
D. Boncioli$^{3}$,
A. di Matteo$^{4}$,
A.F. Grillo$^{3}$,
S. Petrera$^{2,4}$,
F. Salamida $^{5}$
}

\afiliations{
$^1$ INAF Osservatorio Astrofisico di Arcetri, Firenze, Italy \\
$^2$ GSSI, Gran Sasso Science Institute, L'Aquila, Italy \\
$^3$ INFN Laboratori Nazionali del Gran Sasso, Assergi, Italy  \\
$^4$ INFN and Physics Department, University of L'Aquila, L'Aquila, Italy \\
$^5$ Institut de Physique Nucl\'eaire d'Orsay (IPNO), Universit\'e Paris 11, CNRS-IN2P3, Orsay, France \\
}
\email{denise.boncioli@aquila.infn.it}

\abstract{Ultra-High-Energy Cosmic Ray (UHECR) nuclei propagating in cosmological radiation
backgrounds produce secondary particles detectable at Earth. {\it SimProp}
is a one dimensional code for extragalactic propagation of UHECR
nuclei,
inspired by the kinetic approach of Aloisio et al. As in this
approach, only a subset of nuclei and nuclear channels are used as
representative. We discuss the validation of the code and present
applications to UHECR experimental results.
In particular we present the expected
fluxes of neutrinos produced in some astrophysical
scenario.}

\keywords{UHECRs, extragalactic propagation, simulation, neutrino fluxes}

\begin{document}
\maketitle

%Begin a section.
\section{Introduction}

Ultra-High-Energy Cosmic Rays (UHECRs) are observed at extremely 
high energies up to $3\div5 \times 10^{20}$ eV and the determination of their characteristics is of paramount importance 
in unveiling their possible astrophysical sources and/or acceleration processes. One of the key points of their study is related 
to the propagation of UHE particles in intergalactic space. 

The propagation of UHECR from the source to the observer is mainly conditioned by the intervening astrophysical 
backgrounds, such as the Cosmic Microwave Background (CMB) and the Extragalactic Background Light (EBL). 

Several propagation dependent features in the spectrum can be directly connected to the chemical composition of 
UHECR and/or to the distribution of their sources. Among such features, particularly important 
is the Greisen-Zatsepin-Kuzmin (GZK) suppression of the spectrum \cite{bib:GZK}.

In the case of UHE nuclei the flux is expected to show a suppression at the highest energies due to the photo-disintegration 
process on the CMB and EBL fields, with the production of secondary (lighter) nuclei and nucleons: 
$A+\gamma_{CMB,EBL}\to (A-nN)+nN$, where $A$ is the atomic mass number of the nucleus and $n$ the number of emitted nucleons $N$. 
Within this scenario, the energy of the suppression in the spectrum depends on the nuclear species, mainly on their atomic mass number $A$,
and on the details of the astrophysical backgrounds \cite{bib:NucleiAlo}. Particularly relevant is the EBL field, which fixes the energy 
of the onset of the flux suppression \cite{bib:NucleiAlo}.

Another important quantity that, in principle, could affect the flux behavior at the highest energies is the maximum energy $E_{max}$
provided by the sources. In a typical scenario of rigidity dependent acceleration, the maximum energy is related to that of protons through the nucleus charge $Z$, being $E_{max}^{nucl}=Z E_{max}^{p}$. Therefore, for sufficiently low $E_{max}^p$ the UHECR flux steepening at 
the highest energies could be directly linked with the nucleus charge following, in this case, a picture analogous to the
``knee'' behavior observed in the case of galactic CR \cite{bib:disapp}.

From 1960s a flattening has been observed in the UHECR spectrum at an energy around $3\div 6 \times 10^{18}$ eV, which was 
called "the ankle". This feature may be explained in terms of the pair-production dip \cite{bib:dip}, that, like the GZK steepening,
can be directly linked to the interaction of protons with the CMB radiation. The dip arises due to the process of pair production 
suffered by protons interacting with the CMB field $p+\gamma_{CMB}\to p + e^+ + e^-$ \cite{bib:dip}.

If nuclei dominate the UHECR spectrum the ankle could have a different explanation, indicating a transition from a galactic to an extragalactic component.

This contribution describes a Monte Carlo (MC) simulation code, {\it SimProp} \cite{bib:simprop},
developed for the propagation of 
UHE particles (protons and nuclei) through astrophysical backgrounds. In designing such a code, we have focused on a tool which can provide a fast and reliable analysis of the predictions on the spectrum and 
chemical composition.

In its current implementation {\it SimProp} uses a simplified nuclear model and a mono-dimensional 
treatment of the propagation. 
The code (v2r0) was publicily released in May 2012. A new version (v2r1)\footnote{The {\it SimProp} code here presented is 
available for the community upon request to: \textnormal{SimProp-dev@aquila.infn.it}}, whose improvements will be described below, has been recently released.

\section{Main features of the code}

The original computational scheme (v2r0) \cite{bib:simprop} used to propagate charged particles in {\it SimProp} is based on the 
kinetic approach proposed by Aloisio et al. in \cite{bib:NucleiAlo}. The main ingredients of this method are the continuous energy 
loss (CEL) approximation and the assumption of an exact conservation of the particle's Lorentz factor in the 
photo-disintegration process. Neglecting the nucleus recoil in the 
interaction, we can separate the processes that change the Lorentz factor of the particle, 
leaving unchanged the particle type (pair and photo-pion production), from the processes that conserve it, changing the 
particle type (photo-disintegration). 
The CEL approximation consists essentially in neglecting the stochastic behaviour in the interactions of protons with radiation 
backgrounds, which give rise to pair and pion production. Also the pair production energy losses by nuclei are treated analytically and
photopion production by nuclei are neglected. As a further approximation, the representative nuclei are considered stable and neutrons 
and protons are considered as identical particles.

In the process of photo-disintegration of nuclei, the interaction changes the atomic mass $A$. This interaction process is 
simulated by computing, for each interaction channel $i$, the inverse interaction time averaged over the density of the ambient photons:
\begin{equation}
\frac{1}{ \tau_{A,i}(\Gamma)}= \frac{c}{2 \Gamma^2} \int^{\infty}_{\epsilon_{0}(A)} d\epsilon^{'} 
\sigma_{A,i}(\epsilon^{'}) \epsilon^{'} 
\int^{\infty}_{\epsilon^{'}/(2\Gamma)} d\epsilon \frac{n_{\gamma}(\epsilon)}{\epsilon^2}
\label{eq:betadisi} 
\end{equation}
with $\Gamma$ the Lorentz factor of the interacting particle, $\epsilon^{'}$ the 
energy of the background photon in the rest frame of the particle, $\epsilon_{0}(A)$ the threshold of the considered 
reaction in the rest frame of the nucleus $A$, $\sigma$ the relative cross 
section, $\epsilon$ the energy of the photon in the laboratory system and $n_{\gamma}(\epsilon)$ the density of the 
background photons per unit energy. The total inverse interaction time $\tau_A(\Gamma)^{-1}$
can be obtained summing over the all possible photo-disintegration channels $i$. The photo-disintegration cross section 
as well as the relative branching ratios used in this work are taken from \cite{bib:CrossSectionPuget}.

%The dominant channels of photo-disintegration are single and double nucleon emission associated to the Giant Dipole 
%Resonance (GDR) \cite{bib:CrossSectionPuget}. These processes are favored if the energy of the background photon in the 
%rest frame of the nucleus is $\epsilon<30$ MeV. At higher energies in the range $30<\epsilon<150$ MeV a multi-nucleon 
%emission regime takes over, while at energies $\epsilon>150$ MeV the photo-disintegration cross section rapidly goes to 
%zero \cite{bib:CrossSectionPuget}.

Given the approximations described above, the {\it SimProp} computation scheme is a one dimensional algorithm in 
which only the red-shift $z$ follows the "history" of the propagating particle. 
%This approximation together with
%the Lorentz factor conservation in the photo-disintegration process justifies  integrating over the 
%photon density instead of generating the background photon 
%parameters from their distribution.  

Using the default options, the code produces, for each chosen primary nucleus, events drawn from a distribution flat in the logarithm of energy of primaries as well as in the redshift of the sources. Any other physical distribution can be easily obtained by re-weighting a posteriori the simulated data, and appropriately summing the contributions from 
different primaries proportionally to their abundance at generation.  

In order to describe experimental composition observables, simulations are folded with propagation effects in the atmosphere as in \cite{bib:conex,bib:lna}.

\subsection{Validation of the code}
\begin{figure}[t]
  \centering
  \includegraphics[width=0.4\textwidth]{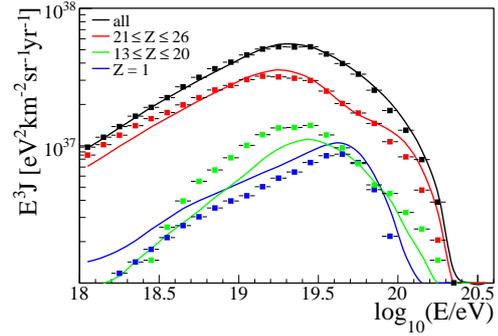}
  \caption{Spectra for pure Iron injection, $\gamma=2.3,~E_{max}(\mathrm{Fe})=5\times10^{21}$ eV \cite{bib:simprop}. Full squares refer to {\it SimProp} simulation while continuous lines refers to CRPropa simulations.}
  \label{f1}
 \end{figure}

The code has been verified against the kinetic approach and compared to other propagation codes. The comparison with respect to CRPropa \cite{bib:crp} is presented in figure \ref{f1}. The differences in the fluxes of the intermediate nuclei can be explained in terms of the set of nuclei and EBL parametrization used in the different codes (\cite{bib:steck,bib:kneiske}). However the effect is negligible on the all-particle spectrum and also for the composition measurements the differences become very small for experimental observables.

\subsection{Applications to selected scenarios}

 \begin{figure}[!t]
  \centering
%  \begin{tabular}{ll}
  \subfigure
  {\includegraphics[width=0.4\textwidth]{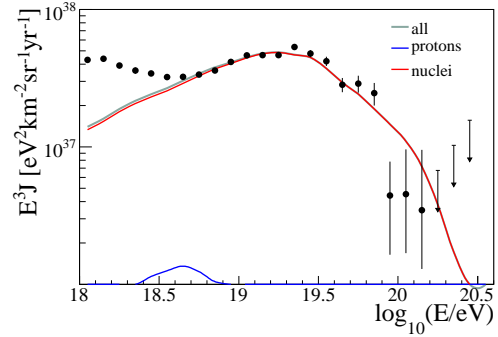}}
\hspace{5mm}  
  \subfigure
  {\includegraphics[width=0.4\textwidth]{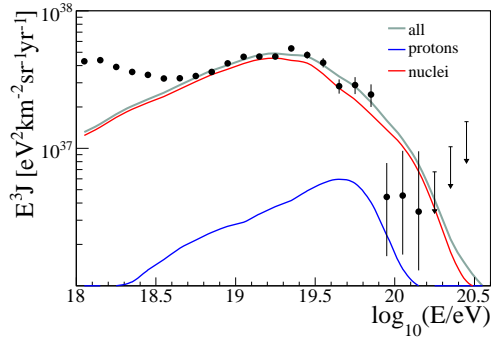} }
 \caption{Spectra for pure Iron injection, $\gamma=2.4$. Upper Panel: $E_{max}(\mathrm{Fe})=10^{20.5}$ eV. Lower Panel: $E_{max}(\mathrm{Fe})=10^{22}$ eV. Red line refers to all secondary nuclei at Earth.}
  \label{f2}
 \end{figure}

Using the simulated data, one can define a {\it source model} identified by:\\
- power spectrum with spectral index $\gamma$ and maximum generation energy $E_{max}$;\\
- abundances of different nuclei at the production.

In general a maximum injection energy $E_{max}$ has been assumed, multiplying the flux by an exponential cut-off function $\exp(1-E/E_{max}(Z))$ for $E > E_{max}$. In the case of a mixed composition a rigidity dependence has been also assumed, i.e. $E_{max}(Z) = (Z/26) E_{max}(Fe)$. However other possibilities can be easily considered.

Each {\it source model} can be compared with experimental data, using both spectra and composition measurements.

In figure \ref{f2} we present results of simulations of  Iron primaries with different maximum production energies, compared to the all particle spectrum obtained by the Auger Observatory \cite{bib:auger}. Although the all particle experimental spectrum can be easily reproduced with a single primary nucleus, this is not the case for the composition measurements \cite{bib:augercomp}. 

\begin{figure}[t]
  \centering
  \includegraphics[width=0.4\textwidth]{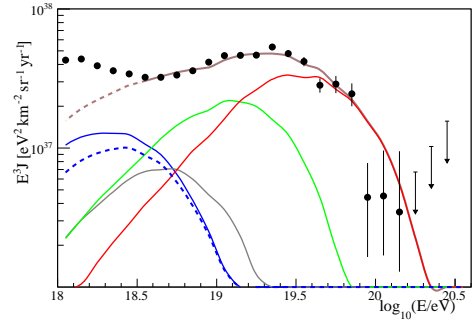}
  \caption{Spectra in the case of source composition in the ``galactic'' inspired scenario (see text). $A = 1$ (blue, dashed line corresponds to primary protons), $2 \leq A \leq 4$ (gray), $9 \leq A \leq 26$ (green) and $27 \leq A \leq 56$ (red), All particle (light brown).}
  \label{f3}
 \end{figure}

\begin{figure}[!t]
   \subfigure
  \centering
  \includegraphics[width=0.4\textwidth]{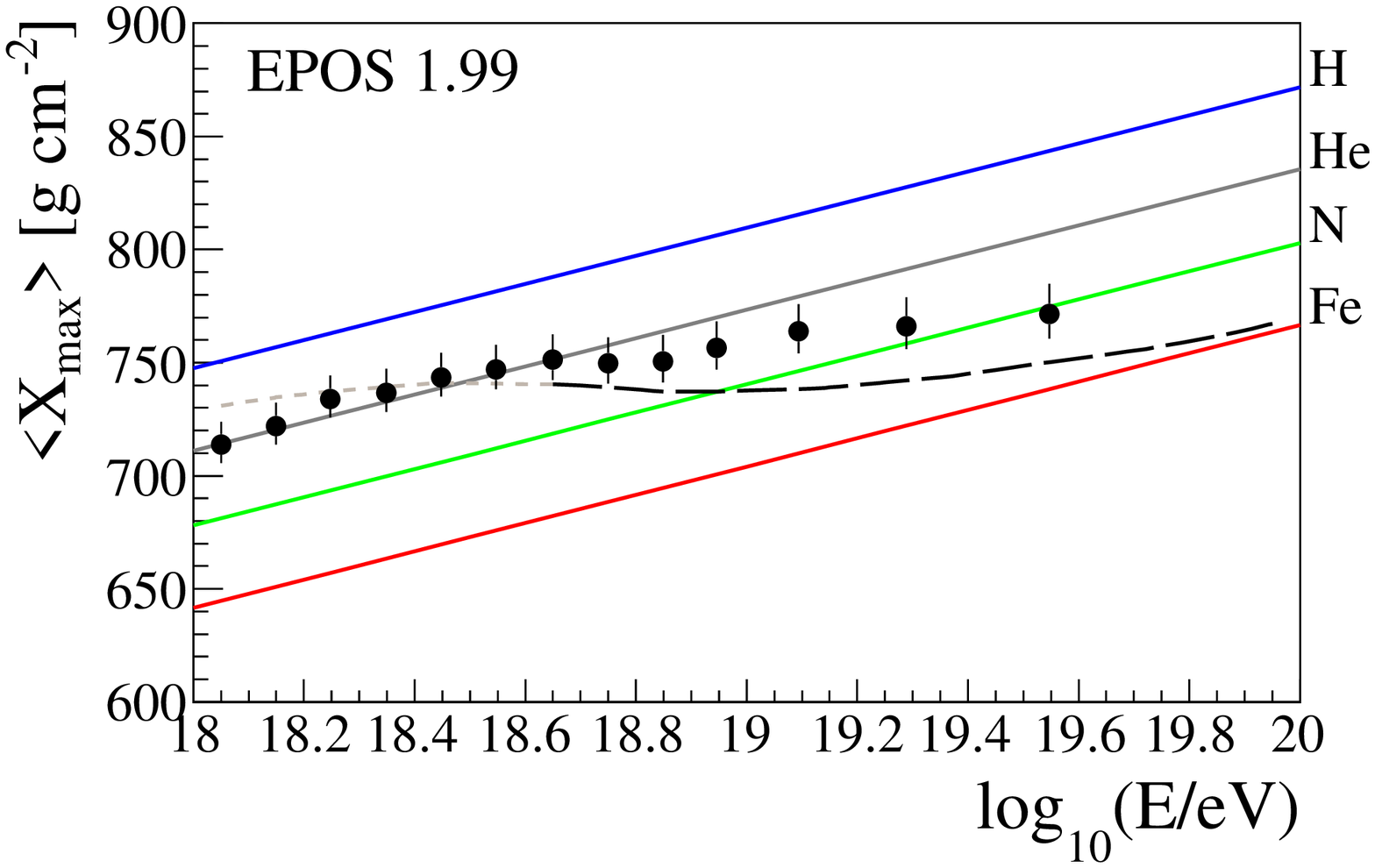}
\subfigure
  \centering
  \includegraphics[width=0.4\textwidth]{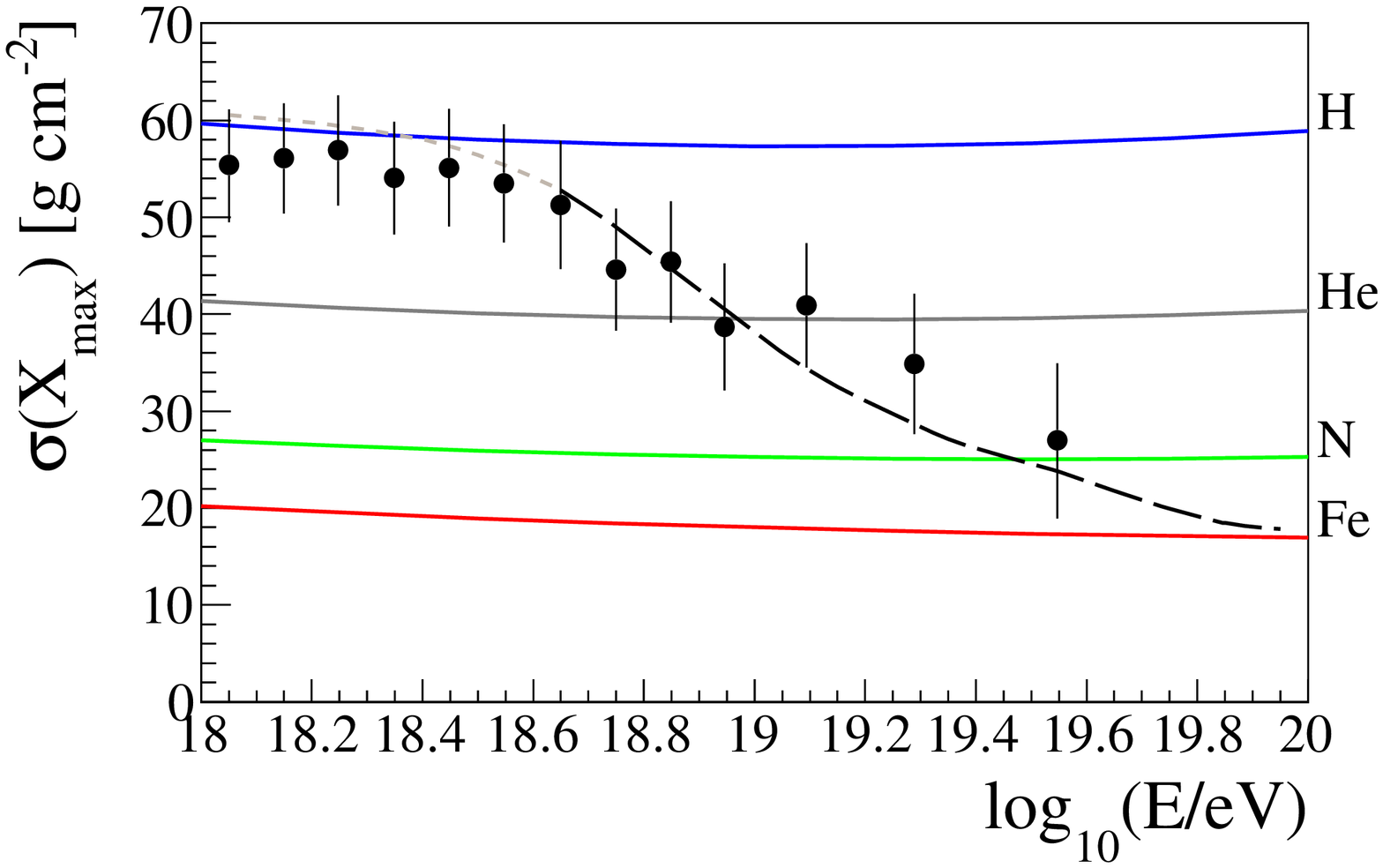}
  \caption{$X_{max}$ (upper panel) and RMS($X_{max}$) in the ``galactic'' inspired scenario.}
  \label{f4}
 \end{figure}

In figures \ref{f3}, \ref{f4} we report the spectrum and composition data obtained in a scenario in which extragalactic sources produce primaries with a composition similar to the measured galactic one \cite{bib:gal}, as assumed in \cite{bib:allard}. To describe the Auger Observatory observations the abundances of elements heavier than Helium are here arbitrarily multiplied by a factor $5$.\\
Figures \ref{f3} and \ref{f4} show that Auger data (spectrum and composition) are reasonably reproduced for the following parameters: $\gamma=1.55$, $E_{max}(\mathrm{Fe}) = 10^{19.9}$ eV.\\
Here the spectrum, as in all figures, is normalized to experimental data for $E \ge 10^{18.6}$ eV. 

\section{The new release}
With respect to the previous version (v2r0), in the new release (v2r1) the fundamental structure has not changed, maintaining the assumption of strict conservation of Lorentz factor and deterministic, continuous energy loss for pair production. 

Pion photoproduction was treated analitically for protons and not implemented for nuclei in v2r0. In v2r1 it is treated as a stochastic,
discrete process both for protons and for nuclei and the decay products of each produced pion are
simulated; this allows to record the redshift (and energies) of the production of neutrinos and photons, which can then be propagated to detection. Photon propagation is not presently implemented; for neutrinos the only energy loss is due to the expansion of the Universe and the correct energy at detection is recorded. Here the main approximations are that only single pion photoproduction has been considered and the pion angular distribution with respect to the parent proton (in the center of mass frame) is always taken to be isotropic.

For what concerns nuclear photodisintegration, in v2r0  all photodisintegration processes result in beta-decay stable isobars and free protons, whether or not this conserves electric charge.
In v2r1, the type of each emitted nucleon is chosen at random, and the
atomic number of the remaining nucleus is chosen so as to conserve
electric charge; any neutrons or unstable nuclei are then assumed to
immediately undergo beta decay (producing neutrinos).

\begin{figure}[t]
  \centering
  \includegraphics[width=0.4\textwidth]{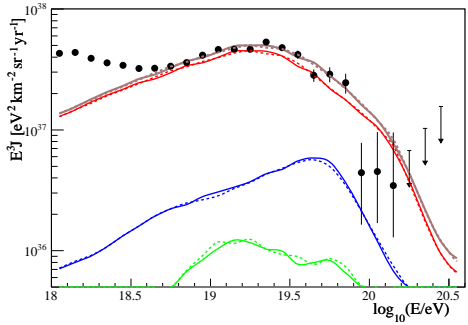}
  \caption{Comparison of spectra obtained with v2r0 (dashed) and v2r1 (continuous line). Injection Iron, $\gamma=2.4,~E_{max}(\mathrm{Fe})=10^{22}$ eV. Colour code as in figure \ref{f3}.}
  \label{f5}
 \end{figure}

The main differences between the versions are connected to the production of pions and their decay. In figure \ref{f5} we present the differences between the two versions in the nuclear sector, in the case of Iron injection, as in figure \ref{f2}, lower panel. 


\section{Neutrino fluxes at Earth}
Given the changes in the code described above, it is now possible to compute neutrino fluxes at 
Earth in various production scenarios. In the following some preliminary results are discussed. (The neutrino flavours listed are those at their production, which cannot be distinguished at Earth due to neutrino oscillations.)

\subsection{ESS (proton-only) model}
Engel, Stecker, and Stanev (ESS) \cite{bib:ESS} computed the fluxes of neutrinos at Earth assuming:\\
- only protons at injection;\\
- energy spectrum $ \propto E^{-\gamma}\exp(-E/E_c)$, where $\gamma = 2$ and $E_c = 10^{21.5}$ eV;\\
- evolution effects for the density of sources (see \cite{bib:ESS}).

\begin{figure}[t]
\subfigure
  \centering
  \includegraphics[width=0.4\textwidth]{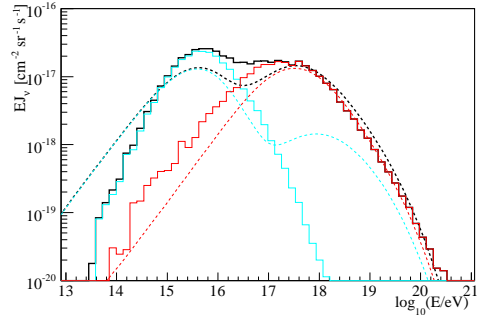}
  \subfigure
  \centering
  \includegraphics[width=0.4\textwidth]{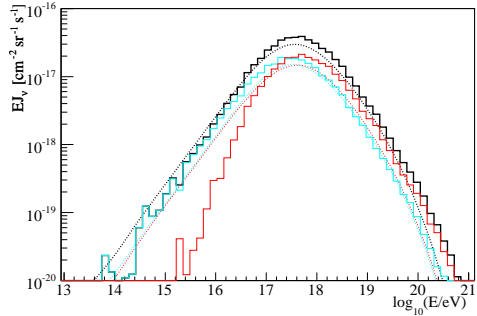}
  \caption{ Fluxes at Earth of $\nu_e$ and $\bar \nu_e$ (top) and $\nu_{\mu}$  and $\bar \nu_{\mu}$(bottom) computed by \textit{SimProp} (solid-line histograms) and ESS (smooth dashed lines): cyan $\bar \nu$; red $\nu$; black sum.}
  \label{f6}
 \end{figure}

Pure proton injection scenarios are at variance with Auger composition measurements \cite{bib:augercomp} but we still computed fluxes under this scenario to compare them with ESS data in order to check the correctness of neutrino production algorithms (figure \ref{f6}). 

The neutrino spectra show a good qualitative agreement with those of ESS, and 
the discrepancies between \textit{SimProp} results and ESS data may be due to approximation we made in treating photohadronic processes, e.g. considering all interactions as single-pion production and as isotropic in the center of mass frame at all energies.

\subsection{The mixed composition model}
\begin{figure}[!t]
  \centering
  \includegraphics[width=0.4\textwidth]{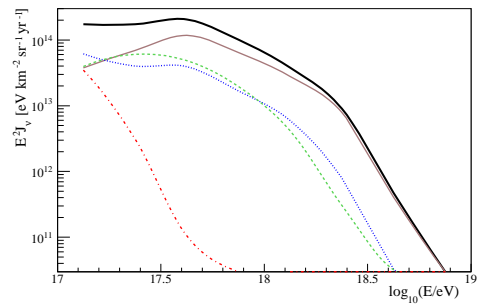}
  \caption{Neutrino spectra in the ``galactic'' inspired scenario (black solid line: total flux; blue dotted: $\nu_e$; red dot-dash: $\bar\nu_e$; green dashed: $\nu_\mu$; brown solid: $\bar\nu_\mu$).} %The $\bar\nu_e$~flux in this energy range is negligible in this scenario.}
  \label{f7}
 \end{figure}

We have then computed the neutrino spectra (see figure \ref{f7}) in the case in which sources produce a mixed composition spectrum with abundances related to the galactic ones, as in section 2.1. 

We find that neutrino fluxes in the ESS scenario are several orders of magnitude larger than in the ``galactic'' scenario. This can be intuitively explained by noticing that the spectrum cut-off in the ESS case, $10^{21.5}$ eV, is much higher than the GZK limit so that a large fraction of the protons can interact. On the contrary in the ``galactic'' case it is about $10^{18.5}(Z/A)$ eV per nucleon, so only the protons and nuclei in  the high energy tail can interact.  If a model with such a low cut-off turns out to be correct, we cannot plausibly expect to detect any UHE neutrinos any time soon in the Auger Observatory.
(On the other hand, photohadronic processes on the EBL --- which are not implemented in \textit{SimProp}~v2r1, but will be in an upcoming version --- are possible at lower nucleon energies than on the CMB, so the neutrino fluxes we computed should be considered lower bounds.)

\section{Conclusions}
In this contribution we have described an extra-galactic propagation code for nuclei {\it SimProp}, developed with the aim of producing a simplified and fast code, yet accurate enough given the experimental uncertainties. A new version of the code has been recently released, that overcomes some limitations of the first one. In particular, the code can now follow neutrinos (and in principle photons) to their detection. 

\vspace*{0.5cm}
\footnotesize{{\bf Acknowledgment:}{We gratefully acknowledge inspiring discussions with our colleagues in the Auger collaboration.}}

\end{document}